\documentclass[a4paper]{article}

\usepackage{INTERSPEECH2021}
\usepackage{cite}
\usepackage{multirow}
\usepackage{todonotes}
\usepackage[
separate-uncertainty = true,
multi-part-units = repeat
]{siunitx}
\usepackage{stfloats}
\usepackage{color,soul}

\title{Cross-lingual Low Resource Speaker Adaptation Using Phonological Features}
\name{Georgia Maniati$^{1\star}$,
	Nikolaos Ellinas$^{1\star}$,
	Konstantinos Markopoulos$^1$,
	Georgios Vamvoukakis$^1$,\\
	June Sig Sung$^2$,
	Hyoungmin Park$^2$,
	Aimilios Chalamandaris$^1$,
	Pirros Tsiakoulis$^1$\thanks{$^{\star}$Equal contribution}}
\address{
	$^1$Innoetics, Samsung Electronics, Greece\\
	$^2$Mobile Communications Business, Samsung Electronics, Republic of Korea}
\email{\{g.maniati, n.ellinas\} @samsung.com}

\begin{document}

\maketitle
\begin{abstract}
The idea of using phonological features instead of phonemes as input to sequence-to-sequence TTS has been recently proposed for zero-shot multilingual speech synthesis. 
This approach is useful for code-switching, as it facilitates the seamless uttering of foreign text embedded in a stream of native text. 
In our work, we train a language-agnostic multispeaker model conditioned on a set of phonologically derived features common across different languages, with the goal of achieving cross-lingual speaker adaptation.
We first experiment with the effect of language phonological similarity on cross-lingual TTS of several source-target language combinations.
Subsequently, we fine-tune the model with very limited data of a new speaker's voice in either a seen or an unseen language, and achieve synthetic speech of equal quality, while preserving the target speaker's identity.
With as few as 32 and 8 utterances of target speaker data, we obtain high speaker similarity scores and naturalness comparable to the corresponding literature.
In the extreme case of only 2 available adaptation utterances, we find that our model behaves as a few-shot learner, as the performance is similar in both the seen and unseen adaptation language scenarios.
\end{abstract}


\section{Introduction}
% inputs for tts
Text-to-speech (TTS) systems have traditionally used sequences of discrete symbols as inputs.
Recently proposed neural architectures \cite{wang2017tacotron,Shen2018} have shown that an efficient end-to-end acoustic model is possible by directly consuming text characters.
The inputs to state-of-the-art TTS systems consist of either text characters (graphemes) or phonemes, with the superiority of phoneme-based systems recently quantified \cite{Fong2019}.
In multilingual TTS, these inputs may originate from various speakers and languages %with different properties or pronunciations, 
introducing variable factors in the model's logic.

Synthesizing speech from multiple speakers with the use of learnable speaker embeddings has been thoroughly examined from the very start of neural TTS \cite{Ping2017} up to most recent efforts \cite{Shen2020}.
Controlling language with learnable embeddings is also straightforward \cite{Zhang2019,Cai2020} and recently, the concept of meta-learning has been shown effective for this purpose \cite{Nekvinda}.
In order to avert the inherent problem of language-dependent speaker representations, domain adaptation has been utilized \cite{Xin2020}.

It is common for the input phoneme representations to be mapped into trainable embeddings, which can be shared across phonemes in the multilingual setting\cite{Zhang2019}.
For this purpose, the International Phonetic Alphabet (IPA) \cite{International1999} can be used \cite{Hemati}.
However, tasks such as code-switching and low-resource language TTS introduce the need for multi-valued representations that will allow the learning of shared qualities across phonemes enabling generalization to previously unseen combinations.

\subsection{Related work}

When investigating multilingual TTS, the input linguistic sequence plays an important role as it incorporates all the distinct language characteristics.
% [again] inputs for tts
Gutkin \textit{et al.} \cite{Gutkin2018, Gutkin2017, Demirsahin2018, Gutkin2017areal} use phonological features (PFs) combining them with phonemes as inputs to multilingual neural TTS models and show improvements in intelligibility across seen and even unseen languages \cite{Gutkin2018}.
Effectively, approaches which concatenate PFs to phonemes do not allow synthesis of unseen phonemes without further training. To this end, Unicode-bytes-based multispeaker multilingual models have been proposed \cite{Li2019, He2021}.
This alternative approach allows unseen characters to be synthesized without entailing any model changes, but since bytes only encode typographical relations, transfer learning of phonological information cannot be achieved for unseen byte combinations.
Staib \textit{et al.} \cite{Staib2020} train a multispeaker variant of Tacotron 2 solely on PFs and show that their model remains unchanged across seen and unseen languages, while enabling the approximation of sounds absent in the training set. %, via transferring learned representations from seen phonemes.
As they aim at code-switched speech, they only train a monolingual and a small multilingual model.
Our work extends the idea of utilizing phonological features to achieve cross-lingual speaker adaptation.
In addition, we present extended experimental results that investigate the effect of language phonological similarity as well as the effect of the adaptation data size.  
Similar features have been used in a feed-forward acoustic model for cross-lingual speaker adaptation using ground truth target phoneme durations \cite{Himawan2020}.

Cross-lingual speaker adaptation can leverage the benefits of a fixed phoneme representation such as IPA.
In \cite{Hemati}, such a model is fine-tuned to the voice of a speaker using 20 minutes of data, while in \cite{Liu2020}, cross-lingual cloning is achieved without further training by utilizing x-vectors extracted from a pretrained system and a common ARPABET phoneme set augmented with stress and tone embeddings.
Language-dependent phones can also be used, as in \cite{Yang2020} where a transformer-based model is trained on a large set of 50 language locales following data imbalance strategies and allowing extensions to new languages with as few as 6 minutes of data.
In the low-resource setting, \cite{DeKorte} apply different language encoders on language-dependent phones and \cite{Luong} show that a learnable linguistic embedding trained in a VAE-like structure can generalize to other languages and adapt to new speakers with few data.

\subsection{Proposed method}

%Proposed Method
In this paper, we apply handcrafted phonological features in cross-lingual TTS and speaker adaptation with very few data.
We follow prior work \cite{lpctron} in order to train a multilingual end-to-end model \cite{Zhang2019} without the addition of language embeddings, since we aim at a model independent of input or output language identity.
First, we investigate if cross-lingual TTS based on phonological features can be improved by using additional training data from typologically related and unrelated languages, and explore the relation of the ratio of unseen phonemes and the perceived intelligibility and naturalness of cross-lingual speech.
Second, we experiment with cross-lingual speaker adaptation and the amount of adaptation data irrespective of the target speakers' native language.
Our results demonstrate that as little as \SI{6}{\second} of adaptation data suffice to achieve synthesized speech highly similar to the target speaker's voice, which is a notable advancement to phoneme-based cross lingual adaptation models requiring \SI{20}{\min} of data for the same task \cite{Hemati}.
The limited data scenario combined with the absence of restrictions to either the target language or the language of the target speaker enables applications such as multilingual TTS of a low-resource language speaker and personalized TTS as long as IPA definitions of the language's phonemes are available.

\section{Method}
\label{sec:method} 

\subsection{Feature set}

%distinctive features
According to PFs' theories, each phoneme of a language can be decomposed into a bundle of simultaneous features.
Jakobson \textit{et al.} \cite{Jakobson1951, Jakobson1956} were the first to introduce a small set of acoustically-defined universal distinctive features. These were later replaced by articulatory features emphasizing their innate nature\cite{Chomsky1968}. 

In our work, features are derived from the articulatory-inspired IPA definitions \cite{International1999}, as categorical multi-valued features. % \cite{Staib2020}. 
%We consider that the acoustic correlates of PFs could be pivotal in transferring learned representations to unseen languages and phonemes in a cross-lingual setting. %, extending the work in \cite{Staib2020}.
%The same categorical multi-valued features as in \cite{Staib2020} are used. 
%That is: \textit{consonant/vowel}; \textit{voicing}, \textit{place}, \textit{manner} and \textit{diacritic} for consonants; \textit{openness}, \textit{frontness} and \textit{roundness} for vowels.
Each feature is encoded into a varied number of dimensions% (e.g. \textit{manner} subsumes values of plosive, nasal, trill, fricative, flap, approximant and lateral)
, resulting in an initial 23-dimensional PFs vector.
The features are 1-hot encoded, except for vowel \textit{openness} and \textit{frontness} which assume continuous values. %according to their IPA definitions. (e.g. close-mid vowels).
All phonemes are split into semi-phonemes before they are fed to the acoustic model to account for cases of changing phoneme quality (diphthongs, affricates).
Each semi-phoneme is mapped to the corresponding PFs vector, resulting in 46 dimensions per phoneme.
For monophthongs, the PFs vector is replicated twice.
Each phoneme is then appended with: a binary \textit{duration} feature differentiating between diphthongs/long vowels/double consonants and regular phonemes; a binary \textit{stress} feature for primary and secondary stress; 7 dimensions for 1-hot encoded punctuation, word boundary, padding and end-of-sequence tokens. 

\subsection{Acoustic model architecture}

The acoustic model follows an attention-based sequence-to-sequence architecture which converts the input linguistic sequence into a sequence of acoustic frames for the LPCNet vocoder \cite{valin2019lpcnet}.
In our case, 
%the input text of any given language is transformed into an IPA phoneme sequence by a TTS front-end module and then each phoneme is mapped to its phonological feature representation by simple dictionary lookup.
the input text of any given language is transformed into internal phonemes by the corresponding frontend module, then mapped to its phonological features’ representation using a dictionary lookup, first to IPA phonemes and then to the PFs vectors.
We leverage the benefits of the reduced dimensionality of the LPCNet features together with a stable Mixture-of-Logistics (MoL) attention module in order to construct a robust model with near natural speech quality \cite{lpctron}.

Since we are working on a multispeaker setup, each speaker is assigned a learnable speaker embedding, which is used to condition the decoder at each step. 
A speaker classifier which predicts the identity of each speaker from the encoder outputs is also used during training as introduced in \cite{Zhang2019}.
This classifier is trained by utilizing the concept of domain adversarial training \cite{Ganin2016} in order to introduce a degree of disentanglement between the linguistic representations and the speaker identity.
The fact that language identity is absent from the model makes this module even more necessary.
Finally, the model is augmented with a residual variational encoder \cite{Hsu2018,Zhang2019} which aims to encode latent factors of audio and increase the naturalness and robustness of the model in the cross-lingual transfer setting.

\subsection{Speaker adaptation}

We want to test the feasibility of adapting the model to an unseen speaker with limited data and enabling them to speak an arbitrary number of languages, regardless of the languages contained in the training set or the speaker's native language.
Our choice of PFs allows the model to be language-independent and as a result there are no restrictions to which speaker or language are compatible with the model without applying any modifications.
%
%After the model is trained, its parameters need to be adapted to the target speaker's voice.
We select a random same-gender speaker-id which is assigned to the target speaker and fine-tune the model for a small number of iterations in a single speaker setting.
During the initial training the encoder has learned meaningful representations of the input PFs and the attention module has learned to align these representations with the acoustic frames.
Since the target speaker's data are limited, we freeze these modules' weights in order to preserve them from forgetting their generic targets. We found this method helps in terms of pronunciation and nativeness of the target language.

\section{Experiments and results}
\label{sec:setup}

\subsection{Data and training setup}

\begin{table}[!t]
	\footnotesize
	\caption{Training dataset details}
	\vspace{-5pt}
	\label{tab:dataset}
	\centering
	\begin{tabular}{@{}llSSS@{}}
		\toprule
		\textbf{Language} & \textbf{Code} & \textbf{Hours} & \textbf{Speakers} & \textbf{Phonemes} \\
		\midrule
		US English & en & 125.9 & 3 & 49  \\
		German     & de & 107.0 & 4 & 58  \\
		French     & fr & 84.7 & 4 & 40  \\
		Spanish    & es & 89.4 & 4 & 33  \\
		Italian    & it & 97.2 & 4 & 57  \\
		Korean     & ko & 164.6 & 3 & 46  \\
		\bottomrule
	\end{tabular}
	\vspace{-10pt}
\end{table}

To train our models, we use an internal multilingual multispeaker dataset comprising 668.8 hours of speech in 6 languages, 22 speakers and 134 unique phonemes. Details about the dataset are shown in Table~\ref{tab:dataset}. 
Varied language configurations' chunks are drawn from this dataset and used as training data throughout our experiments in Sections~\ref{sec:tts} and \ref{sec:adapt}.
%\subsection{Experimental setup}
We use 24 kHz audio data in order to extract the output 22-dimensional acoustic features which consist of 20 Bark-scale cepstral coefficients, the pitch period and the pitch correlation.
The model parameters are trained using the Adam optimizer \cite{Kingma2015}, a batch size of 64 and an initial learning rate of $10^{-3}$, which linearly decays to $3\cdot 10^{-5}$ in 600K iterations.
For speaker adaptation the learning rate is kept stable for another 5K iterations.

%The input text of any given language is transformed into a phoneme sequence using an appropriate TTS front-end module, then each phoneme is mapped to its corresponding PF representation using a 2-step dictionary lookup, i.e. first to IPA and then to the PFs vectors.

\subsection{Formal evaluation}

% Assessment criteria and Metrics
Our models were assessed against naturalness, intelligibility and speaker similarity.
Naturalness was evaluated via mean opinion score (MOS) ranging from 1 to 5, with 1 indicating unnatural and 5 natural speech.
No natural sample of the target language was available for the speakers, as all experiments are cross-lingual, and inserting natural samples of different speakers in the single-speaker tests might affect scores reflecting voice preference.
Speaker similarity evaluation was based on a Likert scale, ranging from ``1: Sounds like a totally different person'' to ``5: Sounds like exactly the same person'' compared to the reference sample.
For intelligibility, listeners were asked to transcribe the generated samples as accurately as possible, and the word error rate (WER) of their responses was computed.

% Listeners
All formal evaluations were conducted online via Amazon Mechanical Turk \cite{Crowston2012}.
Only native speakers of the target language were recruited.
Every audio sample was evaluated by 20 unique participants.
In MOS tests, a validation sample was inserted in each test page to control for potential spurious participants: for naturalness, listeners were instructed to select one response from 1 to 5; for speaker similarity, a different gender voice was used in one of the samples, and the listener was expected to select 1.
After excluding % the responses of 
test pages where participants failed to pass validation% in MOS tests
, and the responses of participants whose WER %in intelligibility 
was over 80\%, the responses of 260 subjects were analyzed for MOS, 174 for intelligibility and 144 for similarity.

% Test corpus
For each test language, 300 sentences were randomly sampled from conversational corpora and Wikipedia articles.
As we were eager to conduct the models' evaluation in the most challenging stress-test setups, 
we subsequently converted the sentences into phonemes and used the corpus selection tool introduced in \cite{Chalamandaris2009}, so as to sort them in descending phonetic coverage order. 
The top 35 phonologically diverse sentences per language comprised each language's test set for all evaluations.

\subsection{Cross-lingual text-to-speech}
\label{sec:tts}

\begin{table}[!t]
	\footnotesize
	\caption{Cross-lingual TTS setup and results of naturalness (MOS with 95\% confidence interval) and intelligibility (WER)}
	\vspace{-5pt}
	\label{tab:results}
	\centering
	\begin{tabular}{@{\hspace*{1mm}} l @{\hspace*{2mm}} l @{\hspace*{2mm}} S @{\hspace*{0mm}} S @{\hspace*{4mm}} S @{\hspace*{1mm}} S @{\hspace*{0mm}} S @{\hspace*{1mm}}}

		\toprule
		\multicolumn{4}{c}{\textbf{Setup}} & \multicolumn{3}{c}{\textbf{Results}}\\
		\midrule
		\textbf{Train} & \textbf{Test} & \multicolumn{2}{c}{\textbf{UPR\%}} & \textbf{WER\%} & \multicolumn{2}{c}{\textbf{MOS}}
		\\
		\midrule
		de    & en & 18.6 & {\scriptsize${\pm}$4.7} & 23.41 & 3.47 & {\scriptsize${\pm}$0.10}   \\
		& fr & 20.3 & {\scriptsize${\pm}$7.4} & 11.97 & 3.23 & {\scriptsize${\pm}$0.11}   \\
		& it & 38.3 & {\scriptsize${\pm}$6.6} & 3.19 & 2.30 & {\scriptsize${\pm}$0.08}   \\
		\midrule
		de+es & en & 11.3 & {\scriptsize${\pm}$3.9} & 15.71 & 3.68 & {\scriptsize${\pm}$0.06}   \\
		& fr &  6.6 & {\scriptsize${\pm}$3.9} & 11.68 & 3.06 & {\scriptsize${\pm}$0.08}   \\
		& it &  7.4 & {\scriptsize${\pm}$4.1} & 4.95 & 2.12 & {\scriptsize${\pm}$0.06}   \\		
		\midrule
		de+es+ko & en & 9.1 & {\scriptsize${\pm}$3.8} & 13.90 & 3.59 & {\scriptsize${\pm}$0.05}  \\
		& fr & 6.6 & {\scriptsize${\pm}$3.9} & 12.11 & 3.09 & {\scriptsize${\pm}$0.06}  \\
		& it & 7.4 & {\scriptsize${\pm}$4.1} & 3.05 & 2.78 & {\scriptsize${\pm}$0.05}  \\
		\midrule
		es+ko   & en & 30.3 & {\scriptsize${\pm}$6.3} & 32.87 & 3.05 & {\scriptsize${\pm}$0.07}  \\
		& fr & 30.4 & {\scriptsize${\pm}$8.5} & 33.27 & 1.70 & {\scriptsize${\pm}$0.06}  \\
		& it & 11.2 & {\scriptsize${\pm}$5.4} & 5.80 & 2.06 & {\scriptsize${\pm}$0.06}  \\
		\midrule
		en+es   & de & 27.1 & {\scriptsize${\pm}$5.9} & 17.77 & 2.48 & {\scriptsize${\pm}$0.09}  \\
		& fr & 15.1 & {\scriptsize${\pm}$6.0} & 15.69 & 2.56 & {\scriptsize${\pm}$0.09}  \\
		& it &  6.9 & {\scriptsize${\pm}$4.1} & 3.10 & 2.72 & {\scriptsize${\pm}$0.06}  \\
		\midrule
		en+fr   & de & 25.8 & {\scriptsize${\pm}$4.7} & 26.86 & 2.21 & {\scriptsize${\pm}$0.10}  \\
		& es &  5.2 & {\scriptsize${\pm}$3.2} & 19.03 & 2.58 & {\scriptsize${\pm}$0.06}  \\
		& it & 13.8 & {\scriptsize${\pm}$3.7} & 6.11 & 2.35 & {\scriptsize${\pm}$0.06}  \\	
		\bottomrule
	\end{tabular}
	\vspace{-10pt}
\end{table}

% setup: train-test combinations & motivation
In our first set of experiments, we attempted cross-lingual text-to-speech; that is, synthesis in languages that are unseen in our models' training data, using combinations of the data described in Table \ref{tab:dataset}.
PFs of unseen phonemes are derived from the IPA.
Due to the nature of our method, the resulting speech is accented and may retain prosodic characteristics of the source languages.
We started with a \textit{de} monolingual multispeaker model, and gradually augmented the training data with \textit{es} and then \textit{ko}, while we evaluated the models' performance in synthesizing \textit{en}, \textit{fr} and \textit{it}. 
As we could only conduct formal evaluations for a limited number of language configurations, in this experiment we opted to investigate whether the addition of data from: (i) a language phylogenetically close to the target language can favour cross-lingual synthesis (the addition of \textit{es} data for \textit{fr} and \textit{it}), (ii) a typologically diverse language (\textit{ko}) can degrade synthesis of a cross-lingual model based on PFs. 
Subsequently, we trained two models on English and one Romance language (\textit{en+es, en+fr}) as well as an \textit{es+ko} model which we aimed to compare to \textit{de+es+ko} with regards to the effect of the proportion of the distanced language data within the training data.

% setup: speakers used for testing
For each training setup, we conducted the experiments using as many test speakers as are the languages in the training setup, and for each language, one female speaker was selected as test voice.
That is, for the \textit{de+es} setup, the \textit{en, fr, it} samples have been generated by a \textit{de} female speaker as well as an \textit{es} female speaker.\footnote{Samples at: https://innoetics.github.io/publications/phonological-features/index.html}
We follow this protocol aiming to control a potential effect of the proximity of the source speaker's language to any of the target languages.
As no significant variation was observed, the results presented in Table \ref{tab:results} are the averaged values among all test speakers. 

% evaluations performed
% Intelligibility (block design)
The intelligibility evaluation of the models was conducted first. 
%Each test comprised samples of a single speaker.
For any given test speaker and target language combination, we created 1-4 intelligibility tests, as many as the number of training setups in which the speaker's data were contained.
Each of the tests comprised the 35 test sentences, where we randomly drew samples from all training setups for the target language, such that each listener would evaluate several models without listening to each sentence more than once.
We conducted minimal processing of the responses prior to analysis, i.e. removed punctuation and normalized case, while we could not resolve homographs or correct spelling mistakes. 
In the second test, we evaluated how natural the cross-lingual samples were.
% setup: UPR between train languages and test sentences
As we are interested in the relationship between the count of unseen phonemes and the target language intelligibility and naturalness, we calculate the unseen phoneme rate (UPR) for each train-test setup. The UPR is computed per test utterance, as the number of phonemes not present in the training language(s), divided by the total number of phonemes in the utterance. We report the mean and standard deviation of the UPR over all test language's utterances.

% results
Table~\ref{tab:results} shows that augmenting the training data of a monolingual or multilingual PFs model with diverse language data can improve intelligibility and naturalness in an unseen language.
This improvement is apparent for \textit{en}, first generated from a monolingual, then a small multilingual and finally an augmented model with \textit{ko}.
Interestingly, the addition of a typologically diverse language in the latter setup improves 
intelligibility, while MOS is not significantly affected.
We observe that when the unrelated language data exceed in quantity the data of the related language, the PFs model's intelligibility is severely affected;
\textit{es+ko}, where the size of \textit{ko} data is almost double the size of \textit{es}, performs poorly in intelligibility in all test languages.
Replacing \textit{ko} with \textit{en} results in a more intelligible model. 
We note that naive listeners assign low MOS scores to accented speech, even if the speech is highly intelligible.
This is the case especially for \textit{it}, where WER is the lowest across all languages but listeners consistently assign low values across setups. 
This could also be linked to cultural factors of perception of accented speech. % as acceptable.
For most target languages, we observe that MOS is negatively correlated to UPR %($r = -0.3$) 
and WER is correlated to UPR%($r = 0.47$)
, suggesting that in high UPR stress test scenarios, the approximation of unseen phonemes cannot attenuate the effects of pronunciation errors in quality.
Our informal evaluation showed that although in most cases the approximations made by the model are to neighbouring phonemes, in \textit{de+es$\rightarrow$it} some phonemes collapse to inappropriate phonemes, affecting results for this model.
Notably, its performance is improved with the addition of the unrelated language data, even if UPR remains unchanged (\textit{de+es+ko} outperforms \textit{de+es} for \textit{it}).

\subsection{Cross-lingual speaker adaptation}
\label{sec:adapt}

\begin{table*}[!t]
	\footnotesize
	\caption{Cross-lingual speaker adaptation setup and MOS results of naturalness and speaker similarity with 95\% confidence interval}
	\vspace{-5pt}
	\label{tab:adaptation}
	\centering
	\begin{tabular}{@{\hspace*{1mm}} l @{\hspace*{3mm}} S @{\hspace*{1mm}} S @{\hspace*{3mm}} S @{\hspace*{0mm}} S @{\hspace*{2mm}} S @{\hspace*{0mm}} S @{\hspace*{2mm}} S @{\hspace*{0mm}} S @{\hspace*{2mm}} S @{\hspace*{0mm}} S @{\hspace*{5mm}} S @{\hspace*{0mm}} S @{\hspace*{2mm}} S @{\hspace*{0mm}} S @{\hspace*{2mm}} S @{\hspace*{0mm}} S @{\hspace*{2mm}} S @{\hspace*{0mm}} S @{\hspace*{1mm}}}
		
		\toprule
		% & & & \multicolumn{16}{c}{\textbf{Results (MOS)}}\\
		%\midrule
		\multicolumn{3}{c}{\textbf{Adaptation Setup}} & \multicolumn{8}{c}{\textbf{Naturalness (MOS)}} & \multicolumn{8}{c}{\textbf{Speaker Similarity (MOS)}} \\
		\midrule
		 & \textbf{utt}& \textbf{dur(\SI{}{\second})} & \multicolumn{2}{c}{\textbf{de}} & \multicolumn{2}{c}{\textbf{fr}} & \multicolumn{2}{c}{\textbf{it}} & \multicolumn{2}{c}{\textbf{es}} & \multicolumn{2}{c}{\textbf{de}} & \multicolumn{2}{c}{\textbf{fr}} & \multicolumn{2}{c}{\textbf{it}} & \multicolumn{2}{c}{\textbf{es}} \\
		 
		\midrule
		 en & 32 & 217 & 3.63 & {\scriptsize${\pm}$0.14} & 3.72 & {\scriptsize${\pm}$0.11} & 3.22 & {\scriptsize${\pm}$0.09} & 3.30 & {\scriptsize${\pm}$0.08} & 4.15 & {\scriptsize${\pm}$0.12} & 4.28 & {\scriptsize${\pm}$0.11} & 3.82 & {\scriptsize${\pm}$0.10} & 4.28 & {\scriptsize${\pm}$0.09} \\
		(seen) & 8 & 41 & 3.78 & {\scriptsize${\pm}$0.15} & 3.84 & {\scriptsize${\pm}$0.11} & 3.38 & {\scriptsize${\pm}$0.08} & 3.38 & {\scriptsize${\pm}$0.07} & 3.99 & {\scriptsize${\pm}$0.14} & 4.22 & {\scriptsize${\pm}$0.12} & 3.83 & {\scriptsize${\pm}$0.10} & 4.13 & {\scriptsize${\pm}$0.09} \\
		& 2 & 6 & 3.25 & {\scriptsize${\pm}$0.18} & 3.11 & {\scriptsize${\pm}$0.13} & 2.88 & {\scriptsize${\pm}$0.10} & 2.89 & {\scriptsize${\pm}$0.09} & 3.60 & {\scriptsize${\pm}$0.17} & 3.68 & {\scriptsize${\pm}$0.14} & 2.99 & {\scriptsize${\pm}$0.11} & 3.44 & {\scriptsize${\pm}$0.12} \\
		
		\midrule
		gr & 32 & 262  & 3.04 & {\scriptsize${\pm}$0.15} & 3.22 & {\scriptsize${\pm}$0.12} & 2.98 & {\scriptsize${\pm}$0.09} & 3.07 & {\scriptsize${\pm}$0.09} & 3.82 & {\scriptsize${\pm}$0.18} & 4.05 & {\scriptsize${\pm}$0.12} & 3.56 & {\scriptsize${\pm}$0.12} & 4.04 & {\scriptsize${\pm}$0.09} \\
		(unseen)& 8 & 78 & 3.45 & {\scriptsize${\pm}$0.13} & 3.54 & {\scriptsize${\pm}$0.12} & 3.17 & {\scriptsize${\pm}$0.09} & 3.14 & {\scriptsize${\pm}$0.09} & 3.84 & {\scriptsize${\pm}$0.17} & 4.21 & {\scriptsize${\pm}$0.11} & 3.63 & {\scriptsize${\pm}$0.12} & 4.13 & {\scriptsize${\pm}$0.09} \\
		& 2 & 20 & 3.20 & {\scriptsize${\pm}$0.13} & 3.22 & {\scriptsize${\pm}$0.12} & 2.97 & {\scriptsize${\pm}$0.09} & 2.91 & {\scriptsize${\pm}$0.09} & 3.80 & {\scriptsize${\pm}$0.17} & 4.13 & {\scriptsize${\pm}$0.12} & 3.48 & {\scriptsize${\pm}$0.12} & 4.00 & {\scriptsize${\pm}$0.10} \\		
		
		\bottomrule
	\end{tabular}
	\vspace{-10pt}
\end{table*}

% training setup
Our PFs model is trained on the entire dataset (Table~\ref{tab:dataset}) for 600K iterations, as diverse language data have been shown to ameliorate its performance. Then, it is fine-tuned to the speaker adaptation data for 5K iterations.
An informal evaluation of adaptation configurations is conducted by alternatively freezing the weights of the attention, the encoder and both.
We conclude that the frozen encoder is key to preserving the pronunciation of the target language, while the frozen attention contributes to the stability of the model, preventing end-of-sentence attention failures. We keep both modules frozen in our adaptation setup.
 
% adaptation data & scenarios
Due to our choice of PFs and the language-independent nature of the model, there are no restrictions to either the target language the model can generate speech in, or to the language of the speaker adaptation data (adaptation language).
%% seen & unseen language
As we aim to investigate how the quality of the adapted speech synthesized by the PFs model is affected by the presence of the adaptation language in the training data, we select
2 male voices from an internal dataset, a native American English speaker (\textit{en}) and a Greek native speaker (\textit{gr}). 
Our corpus selection tool \cite{Chalamandaris2009} was used to sort each voice's corpus on the basis of phonetic coverage of the language, and the 32 most phonologically rich sentences were selected as the adaptation corpus per speaker.
In our first set of experiments, we fine-tune the model using the speaker adaptation data from the seen language (\textit{en}), while in the second, we use the adaptation data from the unseen language (\textit{gr}). 
%% data quantity
Furthermore, within each language setup, we experiment with decreasing the amount of adaptation data.
We are primarily interested in the extent to which we can limit the adaptation data for the multilingual PFs model. 
Also, we are eager to examine whether there is any difference in the quantity of adaptation data required to achieve comparable quality results in case of a seen and an unseen adaptation language.
We formally evaluate models fine-tuned on 32, 8, 2 top utterances, as sorted by the corpus selection tool.
Since these sentences in each of the languages differ in length, and the voices differ in speaking rate, the setups may vary in audio duration.
We formally evaluate all models' performance in test languages unseen in the adaptation data, i.e. \textit{de, fr, it, es}, against naturalness and speaker similarity to the original speaker.
For the similarity tests, the speaker's reference sample is provided in the original language.

% results
%%
As expected, the adapted models presented here perform better than cross-lingual TTS models (Table~\ref{tab:results}) as the test languages are seen during training and thus the output speech is not accented.
%% data quantity
The results in Table~\ref{tab:adaptation} show a tendency of slight deterioration of the quality of the adapted speech with the decrease in the amount of adaptation data, across languages. However, for most languages and setups the differences are not statistically significant and fall within the confidence interval range of the mean, showing that the proposed PFs model is robust to very limited data.
%en
For the seen adaptation language, the decrease from 32 to 8 utterances, i.e to \SI{41}{\second} of speech, bears no significant changes in perceived naturalness or speaker similarity of the adapted speech, across all languages. The extreme \textit{en-2} scenario, i.e. adaptation with \SI{6}{\second} of speech deteriorates the model. 
%gr
For the unseen adaptation language, the same pattern is observed. Notably, \textit{gr-8} appears significantly better in naturalness compared to few setup-test language combinations (\textit{gr-32:de, gr-2:it, gr-2:es}), but given the generalized tendency of \textit{en-8 }and \textit{gr-8} models to perform better, we suspect that models adapted on 8 utterances were favoured by the number of fine-tuning iterations used throughout the adaptation experiments (5K). 
Speaker similarity is retained notably high across data quantity setups without statistically  significant differences.
%comparing both
Comparing the two adaptation language setups, we conclude that the use of an unseen language for adapting the PFs model does not entail more data for achieving comparable results to adapting with a seen language.  

\section{Conclusions}
\label{sec:concl}
%Conclusions . 

In this work, we train a language-agnostic multispeaker Tacotron-based model conditioned on a set of IPA-derived phonological features.
The model can perform cross-lingual TTS in any language and is evaluated in high unseen phoneme rate scenarios of various source-target language configurations. %, producing intelligible speech.
We find that augmenting the training data with diverse language data can improve intelligibility and naturalness in an unseen language. We observe that the cross-lingual speech quality is negatively correlated to the ratio of unseen phonemes.
Subsequently, we show that cross-lingual speaker adaptation with very few data is possible by fine-tuning the model on a new speaker of a seen or unseen language.
We experiment with the effect of the size of the adaptation data on speech quality and find that speaker similarity is retained notably high across data quantity setups.
With as few as 32 and 8 utterances of target speaker data, we achieve high speaker similarity scores and naturalness comparable to similar works.
In the extreme case of 2 utterances, performance is similar for seen and unseen adaptation languages, showing that our PFs model is robust to very limited data.
In future work, we plan to form a better understanding of how the speaker-id assigned to the unseen adaptation language's speaker affects the model. Moreover, we plan to investigate alternative phonological features as well as whether the benefits of such representations can be leveraged for monolingual speaker adaptation in extreme low resource scenarios.

\bibliographystyle{IEEEtran}

\bibliography{mybib,refs}

% \begin{thebibliography}{9}
% \bibitem[1]{Davis80-COP}
%   S.\ B.\ Davis and P.\ Mermelstein,
%   ``Comparison of parametric representation for monosyllabic word recognition in continuously spoken sentences,''
%   \textit{IEEE Transactions on Acoustics, Speech and Signal Processing}, vol.~28, no.~4, pp.~357--366, 1980.
% \bibitem[2]{Rabiner89-ATO}
%   L.\ R.\ Rabiner,
%   ``A tutorial on hidden Markov models and selected applications in speech recognition,''
%   \textit{Proceedings of the IEEE}, vol.~77, no.~2, pp.~257-286, 1989.
% \bibitem[3]{Hastie09-TEO}
%   T.\ Hastie, R.\ Tibshirani, and J.\ Friedman,
%   \textit{The Elements of Statistical Learning -- Data Mining, Inference, and Prediction}.
%   New York: Springer, 2009.
% \bibitem[4]{YourName17-XXX}
%   F.\ Lastname1, F.\ Lastname2, and F.\ Lastname3,
%   ``Title of your INTERSPEECH 2021 publication,''
%   in \textit{Interspeech 2021 -- 20\textsuperscript{th} Annual Conference of the International Speech Communication Association, September 15-19, Graz, Austria, Proceedings, Proceedings}, 2020, pp.~100--104.
% \end{thebibliography}

\end{document}